\documentclass[a4paper]{elsarticle}

\usepackage{amsfonts,amsbsy,amssymb,amsmath}
\usepackage{lineno,hyperref}
\usepackage{graphicx}
\usepackage[table,xcdraw]{xcolor}
\usepackage{float}

\modulolinenumbers[5]

\journal{Chemical Physics Letters}

\biboptions{compress}

\begin{document}

\begin{frontmatter}

\title{Tuning the Branching Ratio in a Symmetric Potential Energy Surface with a Post-Transition State Bifurcation using External Time Dependence}


 \author[label1]{V. J. Garc\'ia-Garrido}
 \author[label2]{M. Katsanikas}
 \author[label2]{M. Agaoglou}
 \author[label2]{S. Wiggins\corref{mycorrespondingauthor}}
 \ead{S.Wiggins@bristol.ac.uk}

 \address[label1]{Departamento de F\'isica y Matem\'aticas, Universidad de Alcal\'a, \\ Alcal\'a de Henares, 28871, Spain.\\[.2cm]}

  \address[label2]{School of Mathematics, University of Bristol, \\ Fry Building, Woodland Road, Bristol, BS8 1UG, United Kingdom.}

 \cortext[mycorrespondingauthor]{Corresponding authors}

\begin{abstract}
Chemical selectivity, as quantified by a branching ratio,  is a phenomenon relevant for many organic chemical reactions. It may be exhibited on a  potential energy surface (PES) that 
features a valley-ridge inflection point (VRI) in the region between two sequential index-1 saddles, with one saddle having higher energy than the other. Reaction occurs when a trajectory crosses the region of the higher energy saddle (the ``entrance channel'') and approaches the lower energy saddle. On both sides of the lower energy saddle, there are two wells and  the question we address is that given an initial ensemble of trajectories, what is the relative fraction of trajectories that enter each well.  For a symmetric PES this fraction is $1:1$. We consider a symmetric PES subject to a time-periodic forcing characterized by an amplitude, frequency, and phase. In this letter we analyse how the branching ratio depends on these three parameters.
\end{abstract}


\end{frontmatter}

\section{Introduction}
\label{sec:intro}

Designing reactions with a specified outcome for the  products is a sought after goal in many areas of reaction dynamics. An approach to this problem in organic chemistry is to choose a reacting molecule whose description via a potential energy surface (PES) displays features that lead to multiple possibilities for the products 
\cite{campos2019designing}. This ``selectivity'', i.e. the distribution of products as quantified by a branching ratio, occurs in many organic reactions and is the topic of numerous reviews \cite{carpenter1998dynamic, ess2008, rehbein2011, hare2017post}. Many studies of selectivity have highlighted the fact that the phenomenon is intimately coupled with the dynamics
\cite{black2012dynamics, yang2018dynamics, lourderaj2008classical, ma2017perspective, nyman2014computational, salem1971narcissistic, wang1973dyanmics, tantillo2010carbocation}. 

The importance of dynamics naturally leads to a study of the behavior of trajectories on a potential energy surface having features that may reveal the mechanisms governing selectivity, and the design of such potential energy surfaces is a topic of current interest \cite{campos2019designing}. A model that has served as a paradigm for studying this topic is a potential energy surface (PES) having two sequential index-1 saddles with no intervening energy minimum. In particular, between the two index-1 saddles, one of higher energy than the other, there is a valley ridge inflection (VRI) point. Reaction occurs when a trajectory crosses the region of the higher e\-nergy saddle (the ``entrance channel'') and approaches the lower energy saddle. On both sides of the lower energy saddle, there are two wells. The question of interest is which well does the trajectory enter (``product selectivity'').

This problem was analyzed in \cite{AGKW2020a} where the two potential wells in the PES were symmetric. For the symmetric PES, the product distributions are known in advance: 50\% go to one well and 50\% go to the other. The main point of that analysis was to understand the phase space structures that enforced the $1:1$ branching ratio. It was expected that this knowledge could then be used to understand how symmetry breaking perturbations in phase space, or external time dependent forcing, could influence the phase space mechanisms that enforce the $1:1$ branching ratio in the symmetric case. This letter makes a contribution to this topic by considering the role of external time dependent forcing on the selectivity.

In order to highlight only the role of the forcing on selectivity, we consider the same symmetric PES as considered in \cite{AGKW2020a}, but subjected to a time-periodic external forcing. The role of the forcing is highlighted in the symmetric case since without the forcing we know a priori that the branching ratio is 1:1. The time-dependent term depends on three parameters: an amplitude ($A$), a frequency ($\omega$), and a phase  ($\phi$), and we consider the effect of each of these parameters on selectivity.

This letter is outlined as follows. In Section \ref{sec:model} we introduce the relevant landscape features of the symmetric PES that determines the two degree-of-freedom (DoF) Hamiltonian model used to understand the emergence of selectivity in these type of chemical systems and we describe the external time-periodic forcing that we apply to this Hamiltonian system. In Section \ref{sec:setup} we describe the design of the numerical calculation that allows us to compute  the branching ratio and  assess the effect of the parameters that define the time-periodic forcing term on selectivity. In Section \ref{sec:results} we discuss the results and in Section \ref{sec:conc} we summarize our findings and discuss further directions for research.

\section{The Time-Dependent Hamiltonian Model}
\label{sec:model}
 
We use the potential energy surface (PES) given in \cite{AGKW2020a}, which is symmetric with respect to the $x$-axis. For this symmetric PES we know that the expected product ratio from the branching of trajectories is $1:1$.

\begin{figure}[htbp]
	\begin{center}
		\includegraphics[scale=0.4]{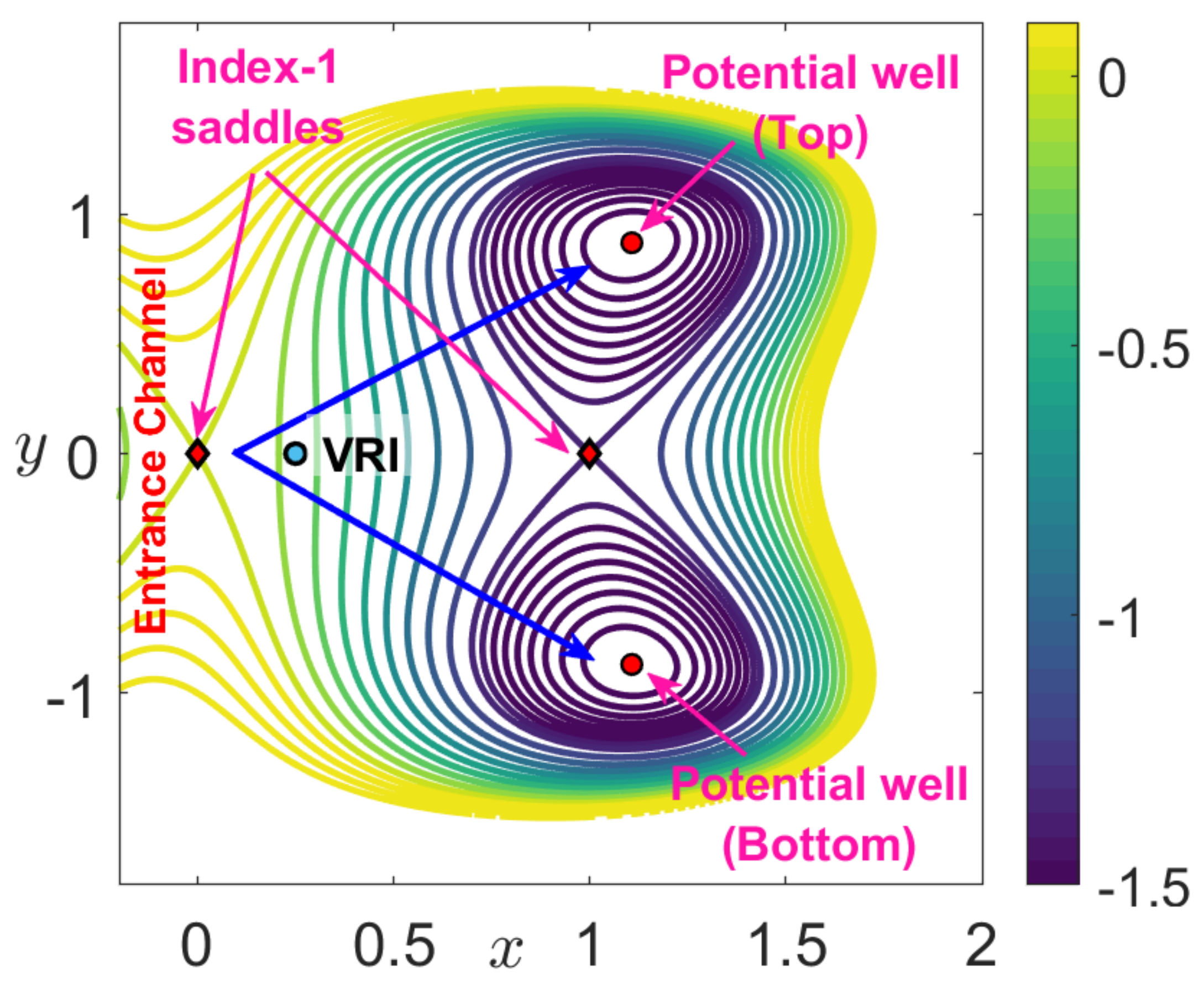}
	\end{center}
	\caption{This figure is from \cite{AGKW2020a}, Equipotential contours of the potential energy surface described in Eq. \eqref{pes_modelVRI}. We denote index-1 saddles (upper and lower) as red diamonds, minima of the potential wells as red points and the valley-ridge inflection point as a blue point. The blue arrows indicate the possible paths for trajectories that enter through the upper channel of the system, illustrating the chemical selectivity mechanism.}
	\label{pes_conts}
\end{figure}

An equipotential of the PES is depicted in Fig. \ref{pes_conts}, which shows an exit/entrance channel that is characterized by an index-1 saddle (upper index-1) and an index-1 saddle (lower index-1) which is an energy barrier separating two potential wells. Moreover, the PES has a valley-ridge inflection point (VRI), which is located between both index-1 saddles. We have also indicated the location of the VRI point and the blue arrows indicate the possible fates of  trajectories that enter the system through the channel of the high energy index-1 saddle. Table \ref{tab:tab1} gives the configuration space coordinates and energies of  the critical points.

\begin{table}[htbp]	
	\begin{tabular}{| l | c | c | c|}
		\hline
		Critical point \hspace{1cm} & \hspace{0.6cm} Location $(x,y)$ \hspace{0.6cm} & \text{Energy} $(V)$ & \hspace{.6cm} \text{Stability} \hspace{.6cm} \\
		\hline\hline
		Index-1 Saddle (Upper) & $(0,0)$ & 0 & saddle $\times$ center \\
		\hline
		Index-1 Saddle (Lower) & $(1,0)$ & -4/3 & saddle $\times$ center \\
		\hline
		Potential Well (Top) & $(1.107146,0.879883)$ & -1.94773 & center \\
		\hline
		Potential Well (Bottom) & $(1.107146,-0.879883)$ & -1.94773 & center  \\
		\hline
		\end{tabular} 
		\caption{Location and energies of the critical points of the PES, together with their linear stability behavior when considered as equilibrium points of Hamilton's equations in the time-independent case, i.e. $A=0$.} 
	\label{tab:tab1} 
\end{table}

The 2 DoF Hamiltonian model that we study is the sum of kinetic energy,  potential energy, and a time-periodic forcing term:

\begin{equation}
H(x,y,p_x,p_y) = \dfrac{p_x^2}{2 m_x} + \dfrac{p_y^2}{2 m_y} + V(x,y) - y A \sin(\omega t + \phi) \;,
\label{hamiltonian}
\end{equation}

\noindent
where $A$, $\omega$, and $\phi$ are variable parameters, $m_{x}$, $m_{y}$ represent the masses of the $x$ and $y$ DoF, respectively.  and the PES has the form:

\begin{equation}
V(x,y) = \dfrac{8}{3}x^3 - 4x^2 + \dfrac{1}{2} y^2 + x y^2 \left(y^2 - 2\right) \;,
\label{pes_modelVRI}
\end{equation}  

\noindent
For this study we choose $m_x = m_y = 1$, and thus Hamilton's equations of motion are the following:

\begin{equation}
\begin{cases}
\dot{x} = \dfrac{\partial H}{\partial p_x} = p_x \\[.4cm]
\dot{y} = \dfrac{\partial H}{\partial p_y} = p_y  \\[.4cm]
\dot{p}_x = -\dfrac{\partial H}{\partial x} = 8 x \left(1 - x\right) + y^2\left(2 - y^2 \right) \\[.4cm]
\dot{p}_y = -\dfrac{\partial H}{\partial y} = y \left[4 x \left(1 - y^2\right) - 1\right] + A \sin(\omega t + \phi)
\end{cases}
\;.
\label{ham_eqs}
\end{equation}

\section{Set-up for the Calculation}
\label{sec:setup}

In this section we describe the setup for our numerical calculations. We emphasize that for a time-dependent Hamiltonian total energy is not conserved. However, we will use energy surface information  for the time independent Hamiltonian system, i.e. $A=0$, with total energy $H = 0.1$ (the same as considered in \cite{AGKW2020a}). 

\begin{figure}[htbp]
	\begin{center}
		\includegraphics[scale=0.35]{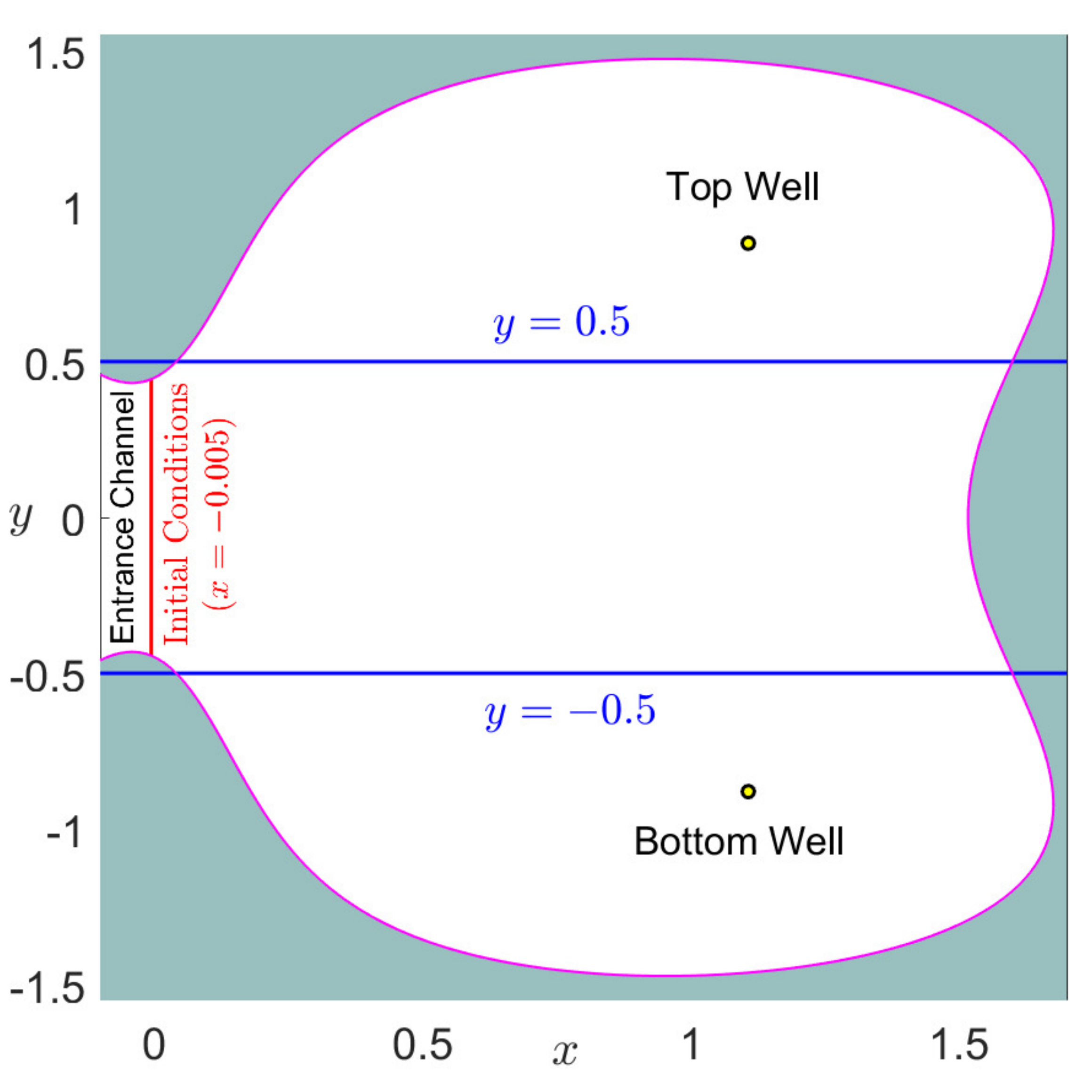}
	\end{center}
	\caption{Configuration space representation of the different elements used to setup the numerical experiment for the calculation  of the branching ratio. We have indicated the locations of the top and bottom wells with yellow dots,  and the red line represents the ensemble of initial conditions used in the simulation. The blue lines at $y = \pm 0.5$ establish the conditions used to determine that a given trajectory has entered the top or bottom well, respectively.}
	\label{fig:experiment_setup}
\end{figure}

In Fig. \ref{fig:experiment_setup} we describe the main components of phase space and configuration space that we use to initiate our numerical experiment, as well as to interpret the results of the numerical experiment. The red line $x = -0.005$ in configuration space shown in the figure  is used to construct the dividing surface on which we initialize the trajectories. One thousand points are equally spaced along this line. For each point we choose the momentum $(p_x, p_y) = (p_x>0, 0)$ such that the total energy of the time-independent Hamiltonian is $H = 0.1$. The condition $p_x > 0$ corresponds to trajectories that enter that react. These are the initial conditions that are then evolved under the time dependent dynamics \eqref{ham_eqs}. The resulting reacting trajectories can have three possible fates for the length of integration time that we consider. They can enter the top well, the bottom well, or they can reflect off the right hand wall of the potential energy surface without entering either well and exit the reacting region through a neighborhood of the high energy index-1 saddle. 

In Fig. \ref{fig:experiment_setup} we have indicated the locations of the top and bottom wells with yellow dots and the red line represents the ensemble of initial conditions used in the simulation.  The blue lines in the figure  at $y = \pm 0.5$ establish the conditions used to determine that a given trajectory has entered the top or bottom well, respectively. When the $y$ coordinate of a trajectory exceeds $y=0.5$ we consider it to have entered to top well, and the integration of that trajectory is stopped, and  when the $y$ coordinate of a trajectory decreases below $y=-0.5$ we consider it to have entered to bottom well, and the integration of that trajectory is stopped.

\section{Results}
\label{sec:results}

In this section we discuss the results of our calculations. We consider first the behavior of the system when it is subjected to an external forcing, for which we use three different values of the angular frequency $\omega = \frac{\pi}{3},\frac{\pi}{2},\pi$. The simulations are carried out in the parameter space determined by the amplitude ($A$) and phase $(\phi)$ of the wave, and the values taken for these parameters move in the ranges $0 \le A \le 0.35$ and $0 \le \phi \le \pi$. 

\begin{figure}[htbp]
	\begin{center}
		A)\includegraphics[scale=0.26]{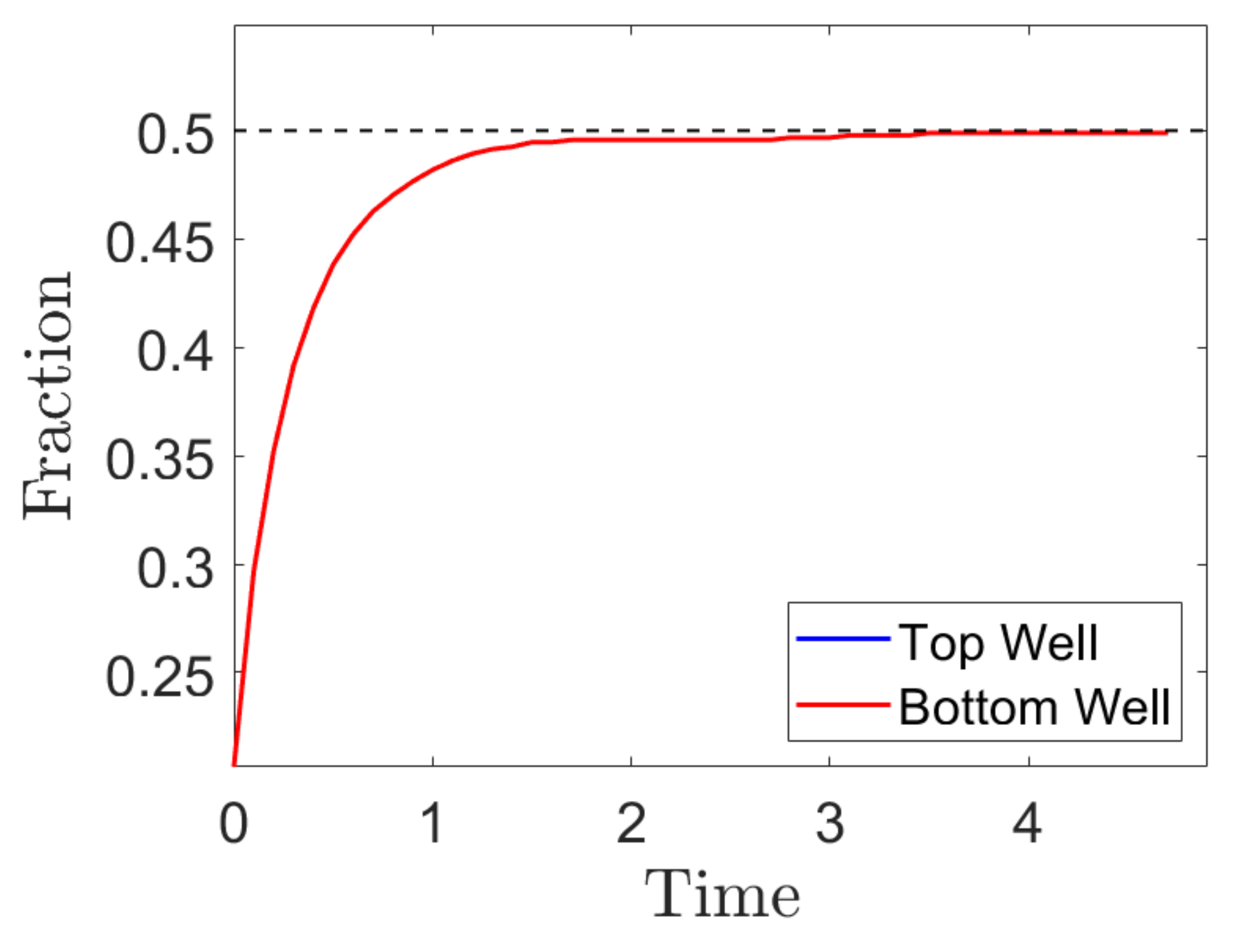}
		B)\includegraphics[scale=0.26]{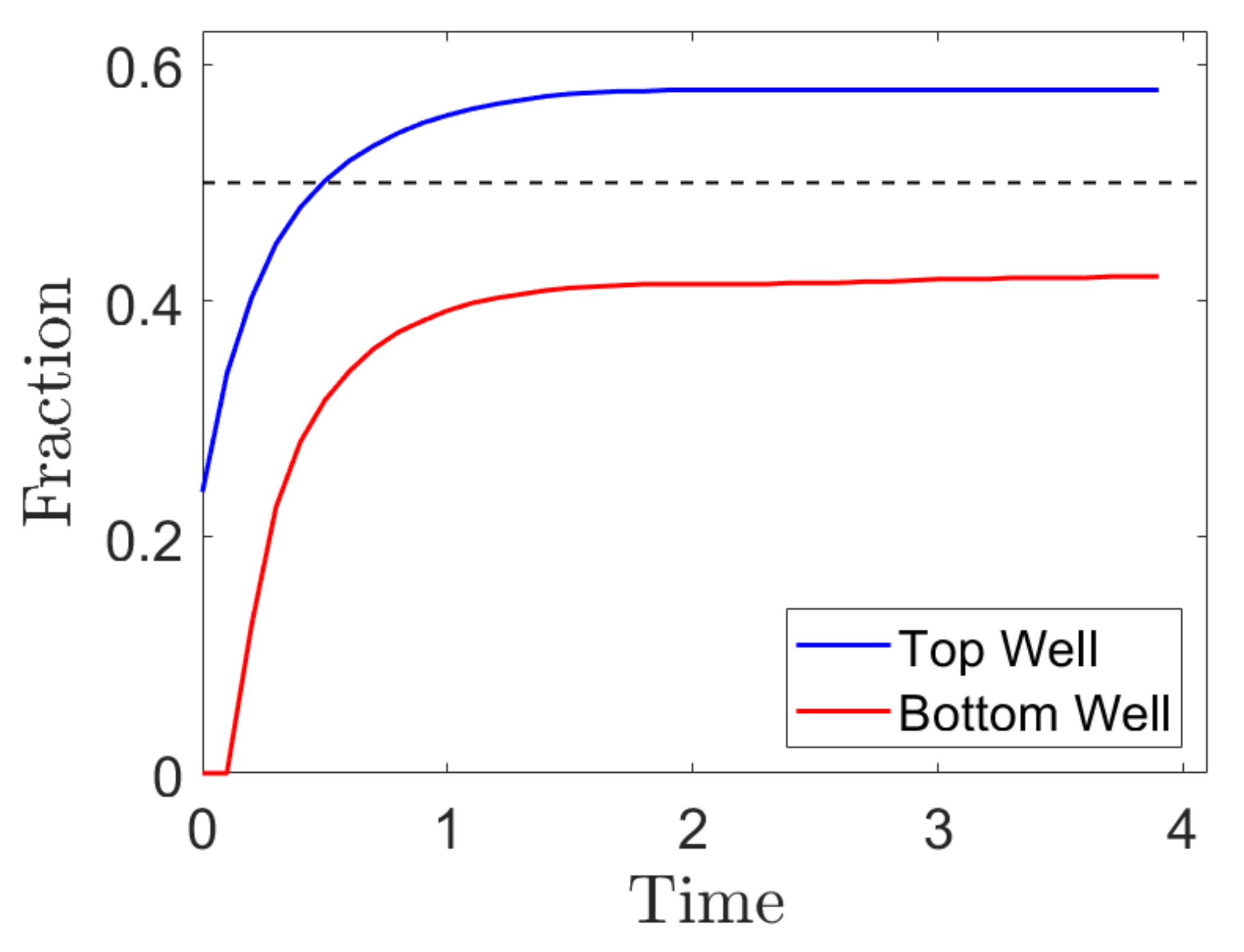}
	\end{center}
	\caption{Time evolution of the fraction of trajectories that enter the top or bottom wells, measured from the total number of trajectories that visit any of the wells. A) Time-independent system, $A=0$. B) Time-dependent system with a time dependence characterized by a forcing with amplitude $A = 0.1$, phase $\phi = 0$ and angular frequency $\omega = \pi / 3$.}
	\label{fig:top_bot_fraction}
\end{figure}

We begin by illustrating the main effect that time dependence can have on the branching ratio. In Fig. \ref{fig:top_bot_fraction} A we show the total fraction of trajectories that enter the top well  (blue curve) and the bottom  well (red curve) as a function of time for $A=0$, i.e.  for the symmetric, time independent PES.  The result is as we expect for the symmetric case. The two curves lie on top of each other. In Fig. \ref{fig:top_bot_fraction} B we consider the case $A=0.1$, $\phi =0$, and $\omega = \frac{\pi}{3}$. In this case we see that the red and blue curves split, and that the branching ratio is no longer $1:1$. We now want to explore the parameter dependence of this effect more completely.

\begin{figure}[htbp]
	\begin{center}
		A)\includegraphics[scale=0.19]{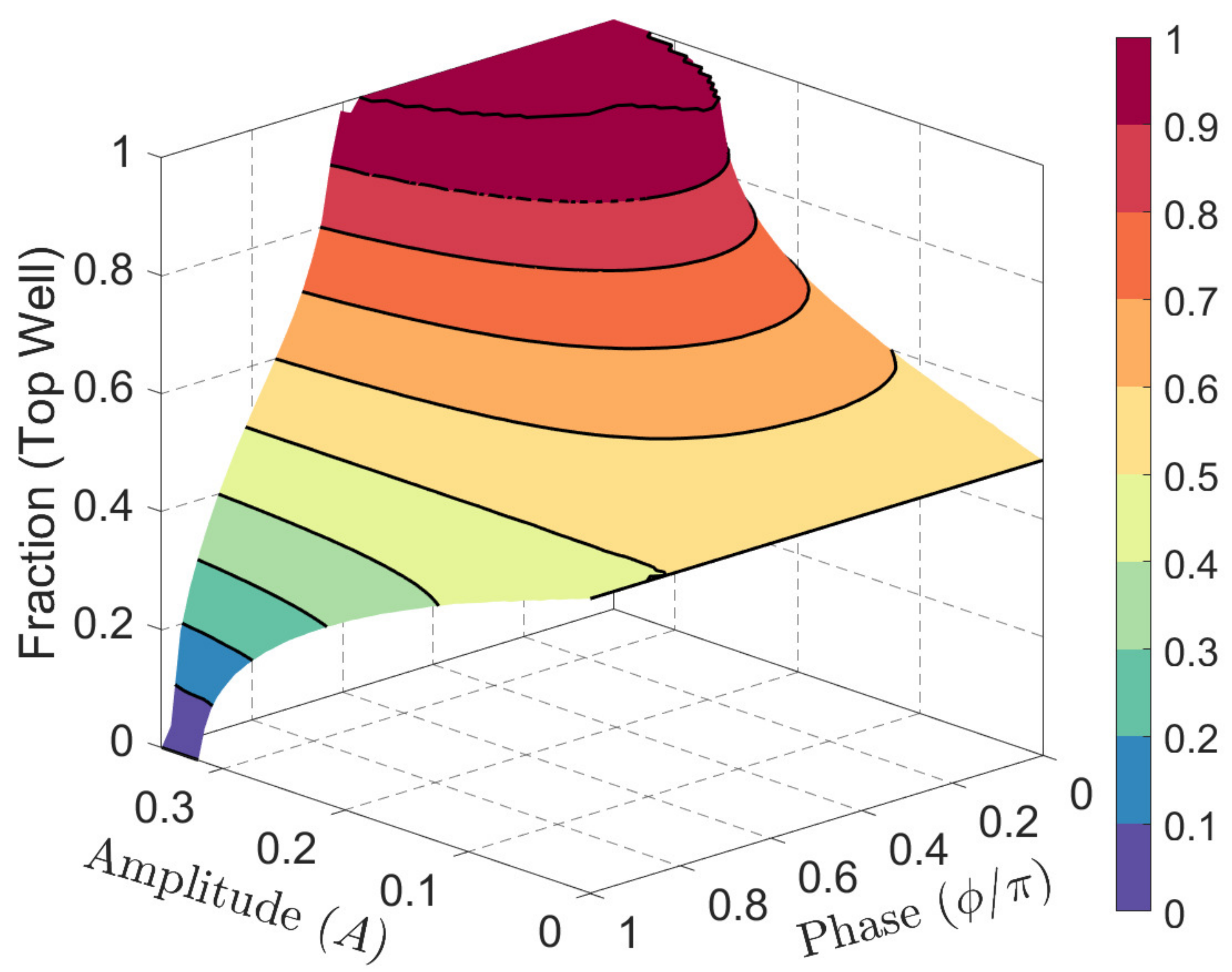}
		B)\includegraphics[scale=0.19]{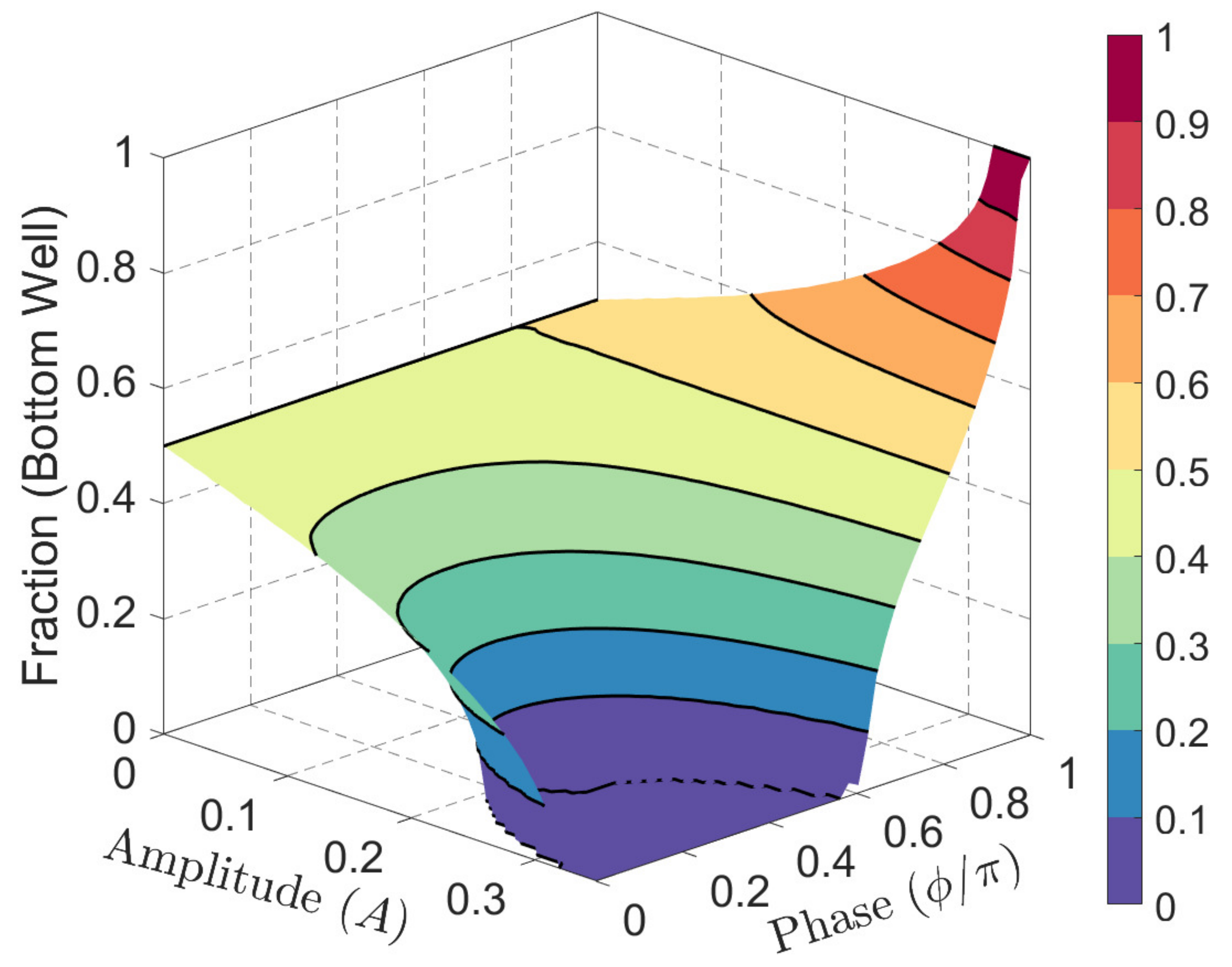}
		C)\includegraphics[scale=0.22]{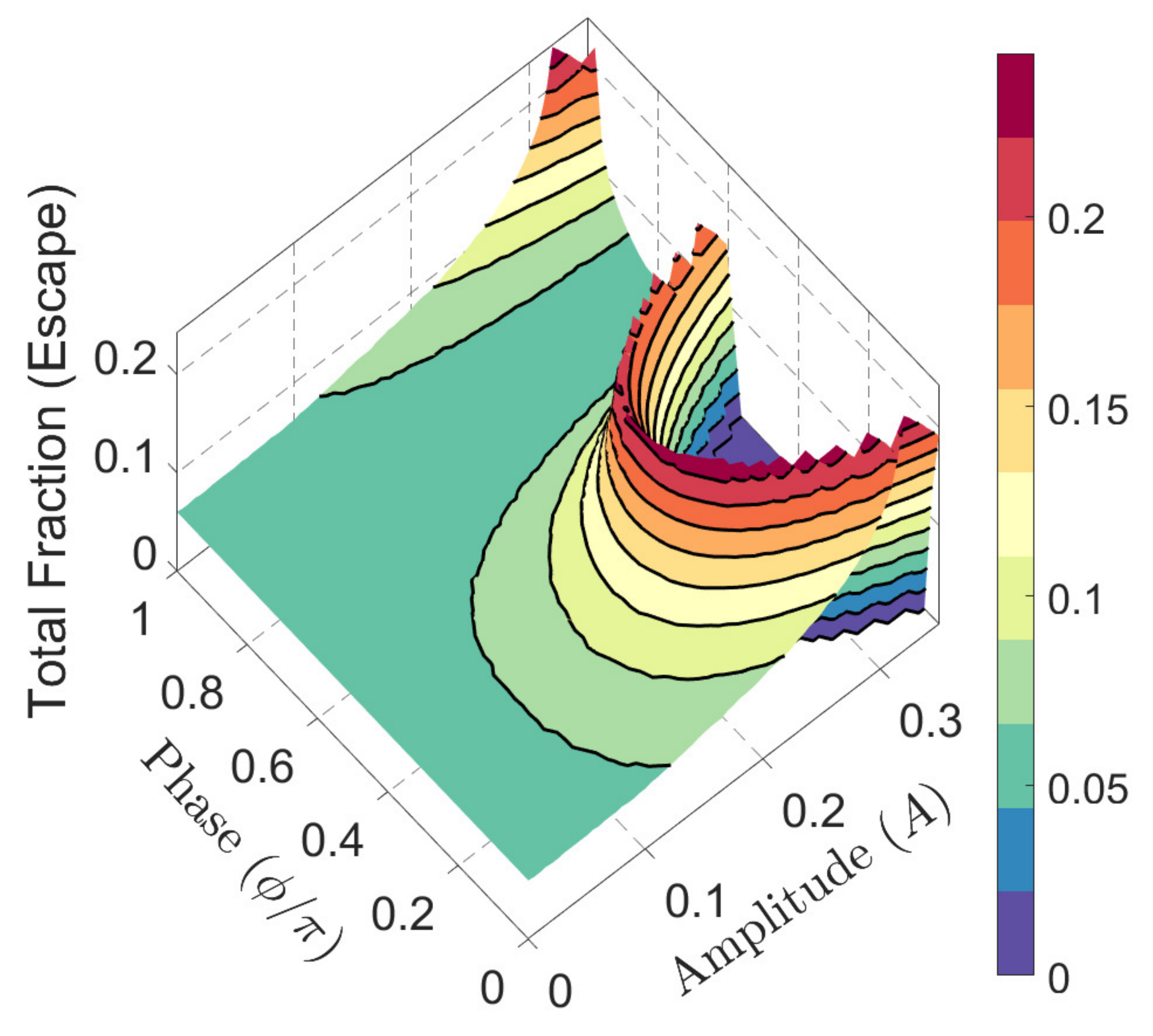}
	\end{center}
	\caption{Fraction of trajectories escaping the system or that enter either of the wells. The amplitude-phase parameter space has been calculated for a time dependent forcing characterized by an angular frequency $\omega = \pi / 3$. A) Trajectories that enter the top well from all those that enter any of the wells. B) The same but for the bottom well. In the panels A and B we calculated the fraction of the trajectories from all those that enter any of the wells. C) Trajectories that escape the system without entering either of the wells, calculated from the total number of trajectories of the initial ensemble.}
	\label{fig:paramSp_pi_div_3}
\end{figure}

\begin{figure}[htbp]
	\begin{center}
		A)\includegraphics[scale=0.18]{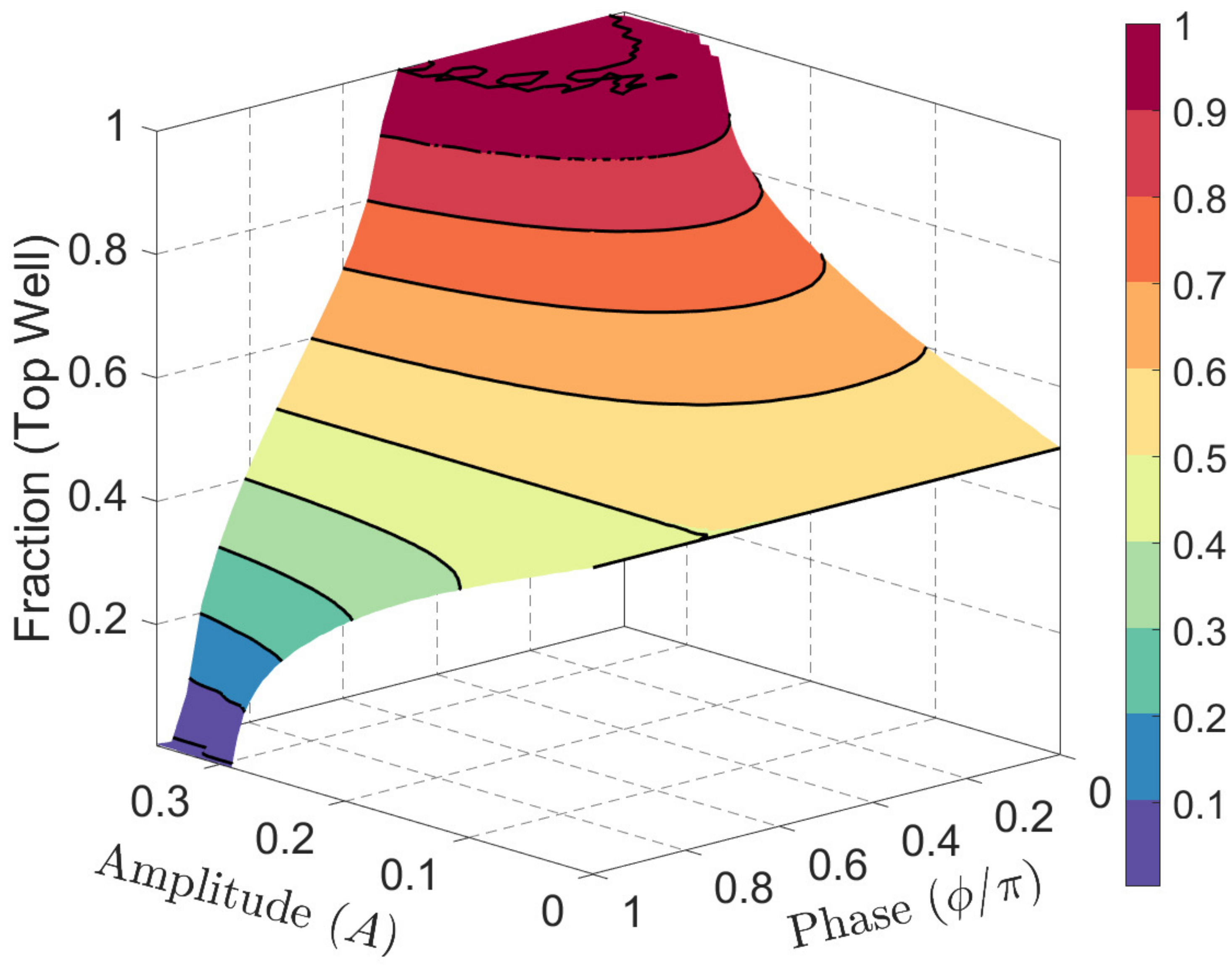}
		B)\includegraphics[scale=0.18]{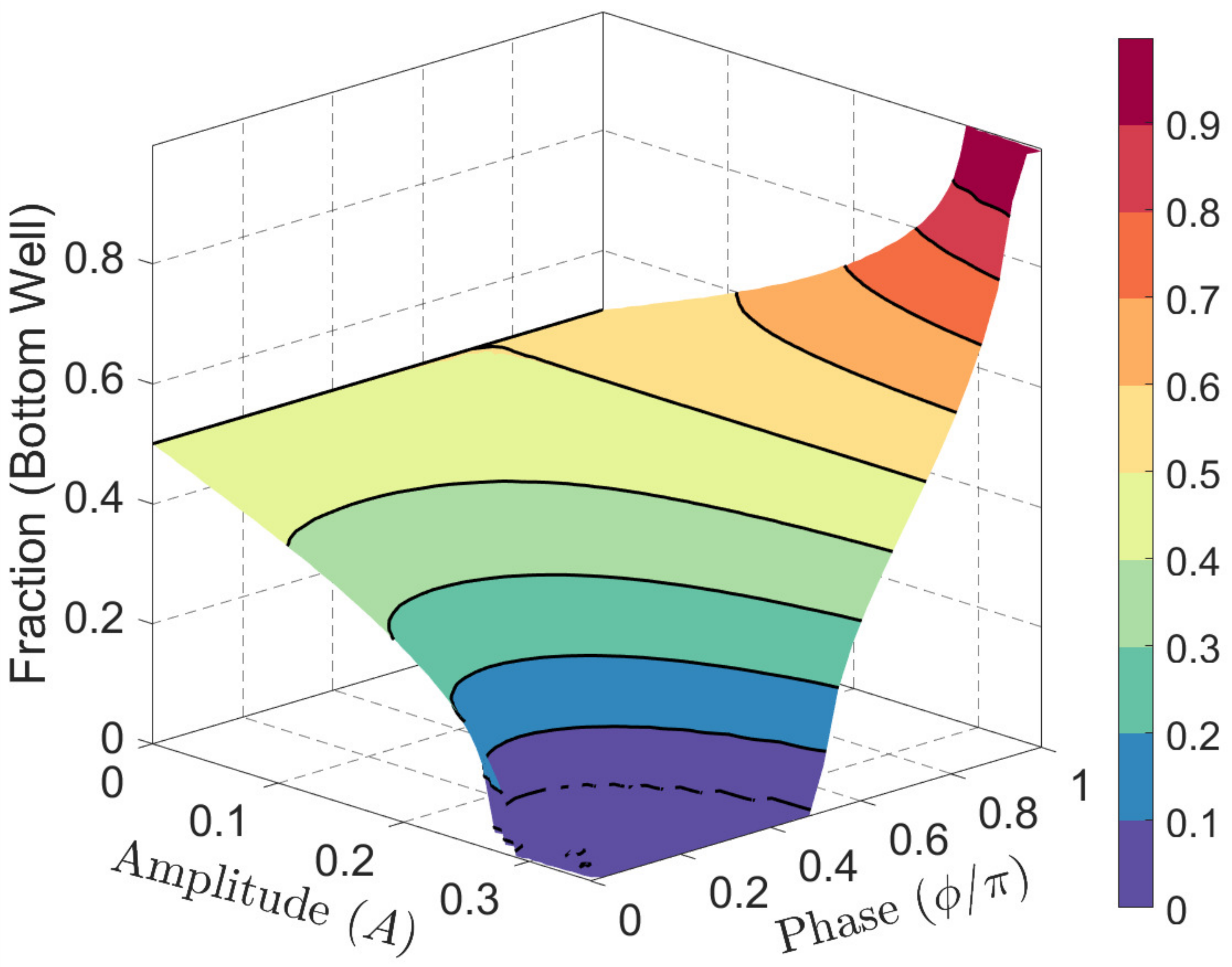}
		C)\includegraphics[scale=0.22]{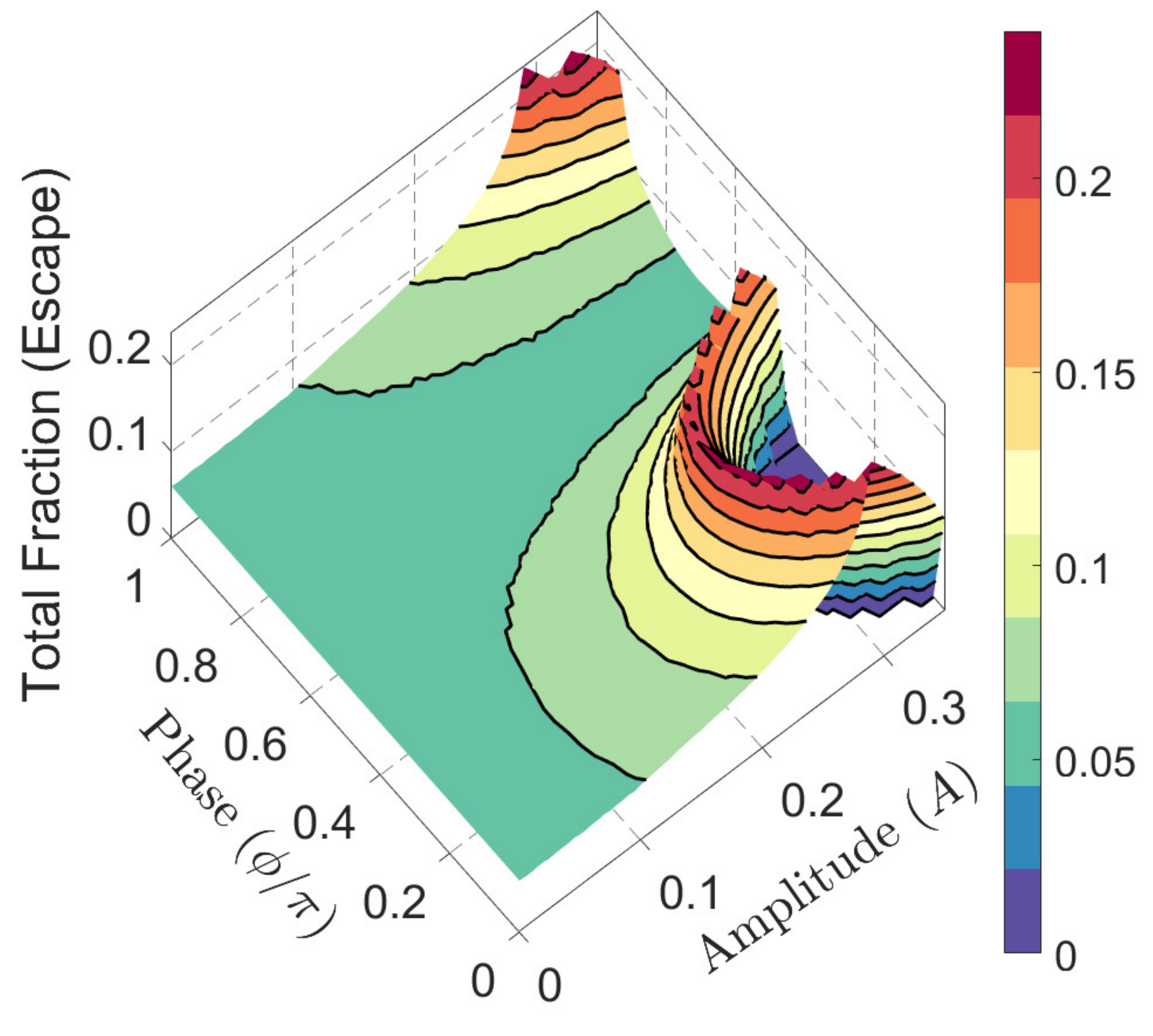}
	\end{center}
	\caption{Fraction of trajectories escaping the system or that enter either of the wells. The amplitude-phase parameter space has been calculated for a time dependent forcing characterized by an angular frequency $\omega = \pi / 2$. A) Trajectories that enter the top well from all those that enter any of the wells. B) The same but for the bottom well. In the panels A and B we calculated the fraction of the trajectories from all those that enter any of the wells. C) Trajectories that escape the system without entering either of the wells, calculated from the total number of trajectories of the initial ensemble.}
	\label{fig:paramSp_pi_div_2}
\end{figure}

\begin{figure}[htbp]
	\begin{center}
		A)\includegraphics[scale=0.19]{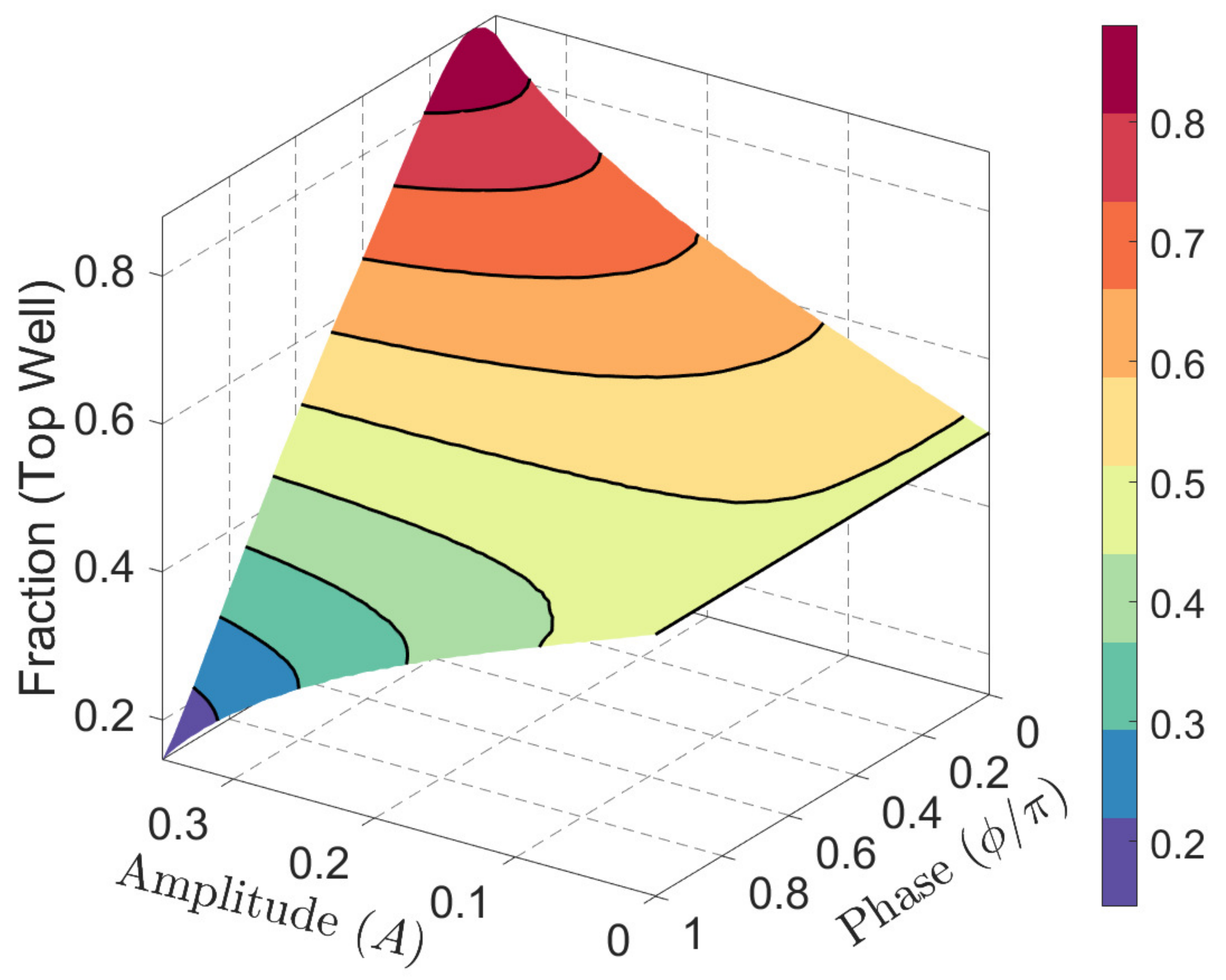}
		B)\includegraphics[scale=0.19]{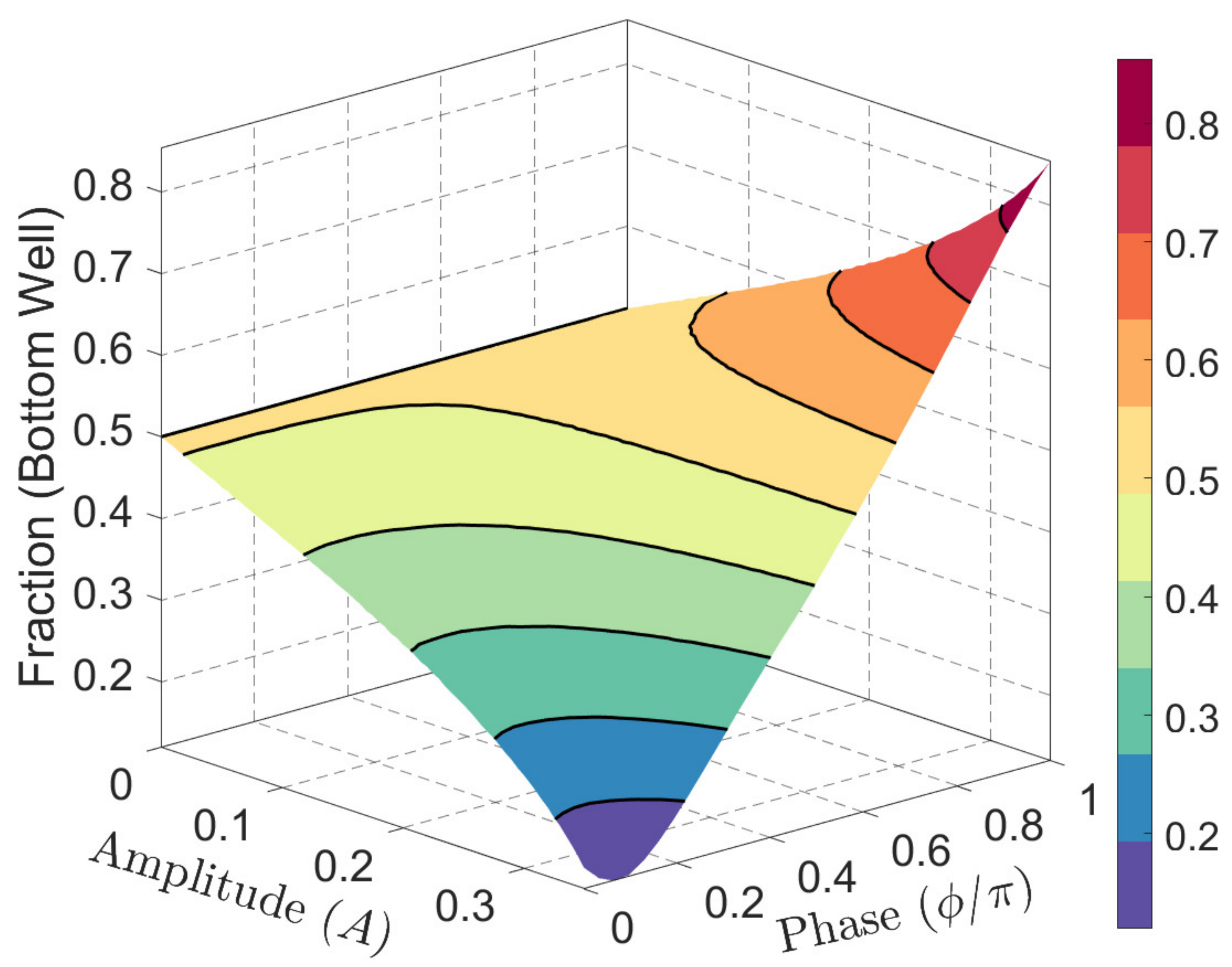}
		C)\includegraphics[scale=0.22]{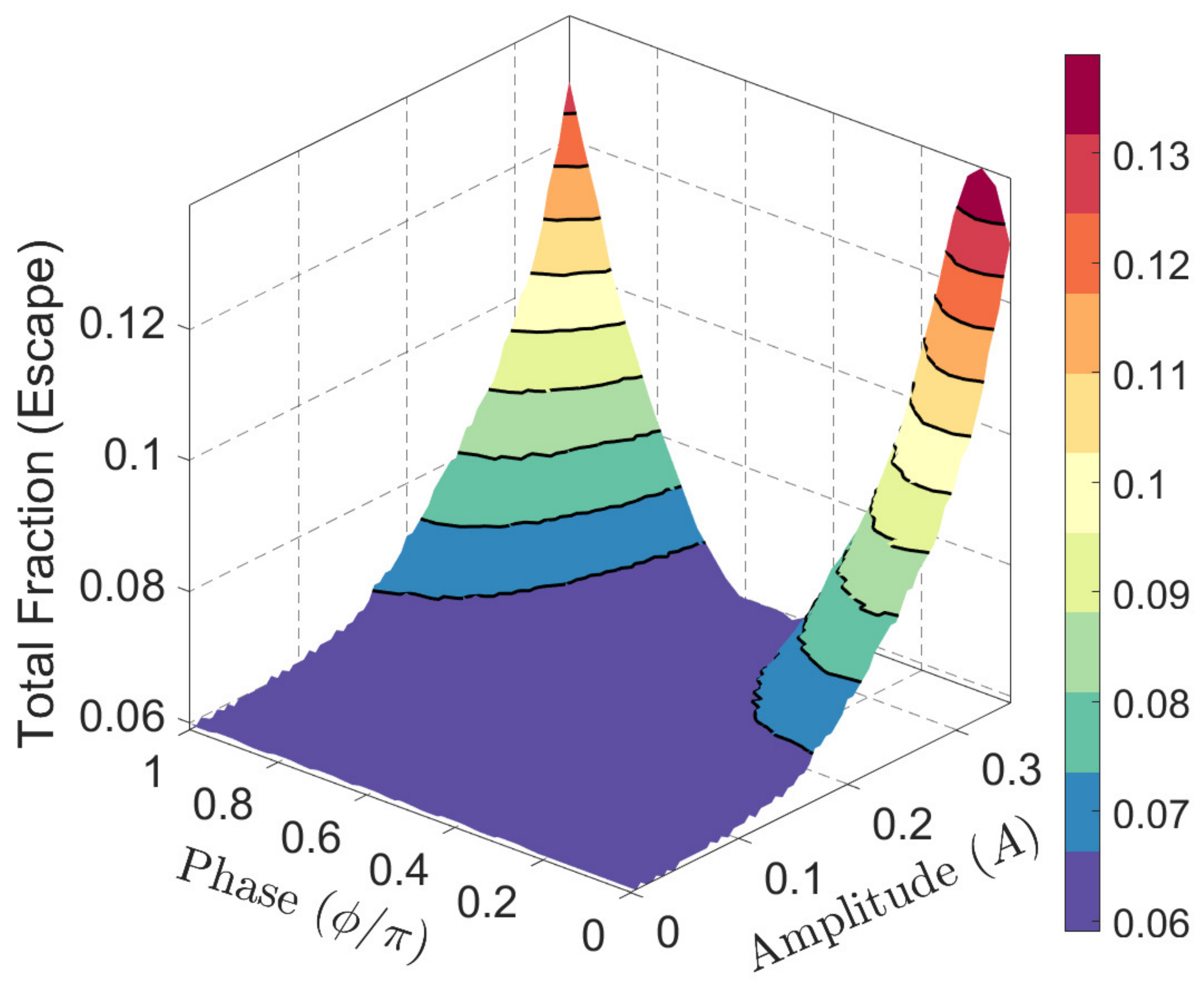}
	\end{center}
	\caption{Fraction of trajectories escaping the system or that enter either of the wells. The amplitude-phase parameter space has been calculated for a time dependent forcing characterized by an angular frequency $\omega = \pi$. A) Trajectories that enter the top well from all those that enter any of the wells. B) The same but for the bottom well. In the panels A and B we calculated the fraction of the trajectories from all those that enter any of the wells. C) Trajectories that escape the system without entering either of the wells, calculated from the total number of trajectories of the initial ensemble.}
	\label{fig:paramSp_pi}
\end{figure}

Figures \ref{fig:paramSp_pi_div_3}, \ref{fig:paramSp_pi_div_2} and \ref{fig:paramSp_pi} correspond respectively to $\omega = \frac{\pi}{3}$, $\omega = \frac{\pi}{2}$ and $\omega = \pi$. For each figure, panel A gives the fraction of trajectories that enter the top well, panel B gives the fraction of trajectories that enter the bottom well, and panel C gives the total fraction of trajectories that do not interact with either well. Notice that Figs. \ref{fig:paramSp_pi_div_3} and \ref{fig:paramSp_pi_div_2} show similar trends in all three panels. For a range of phases (roughly, $0 \le \phi / \pi \le 0.6$)  the fraction of trajectories  in the top well increases as A increases and  for a range of phases (roughly, $0.6 \le \phi / \pi \le 1$)   the fraction of trajectories in the bottom well increases as $A$ increases. Note that essentially all reacting trajectories enter either the top or bottom well. The behaviour in Fig. \ref{fig:paramSp_pi} is somewhat different. For this larger frequency a significant fraction of the trajectories do not either  well before returning to cross the higher energy saddle.

\begin{figure}[htbp]
	\begin{center}
		A)\includegraphics[scale=0.18]{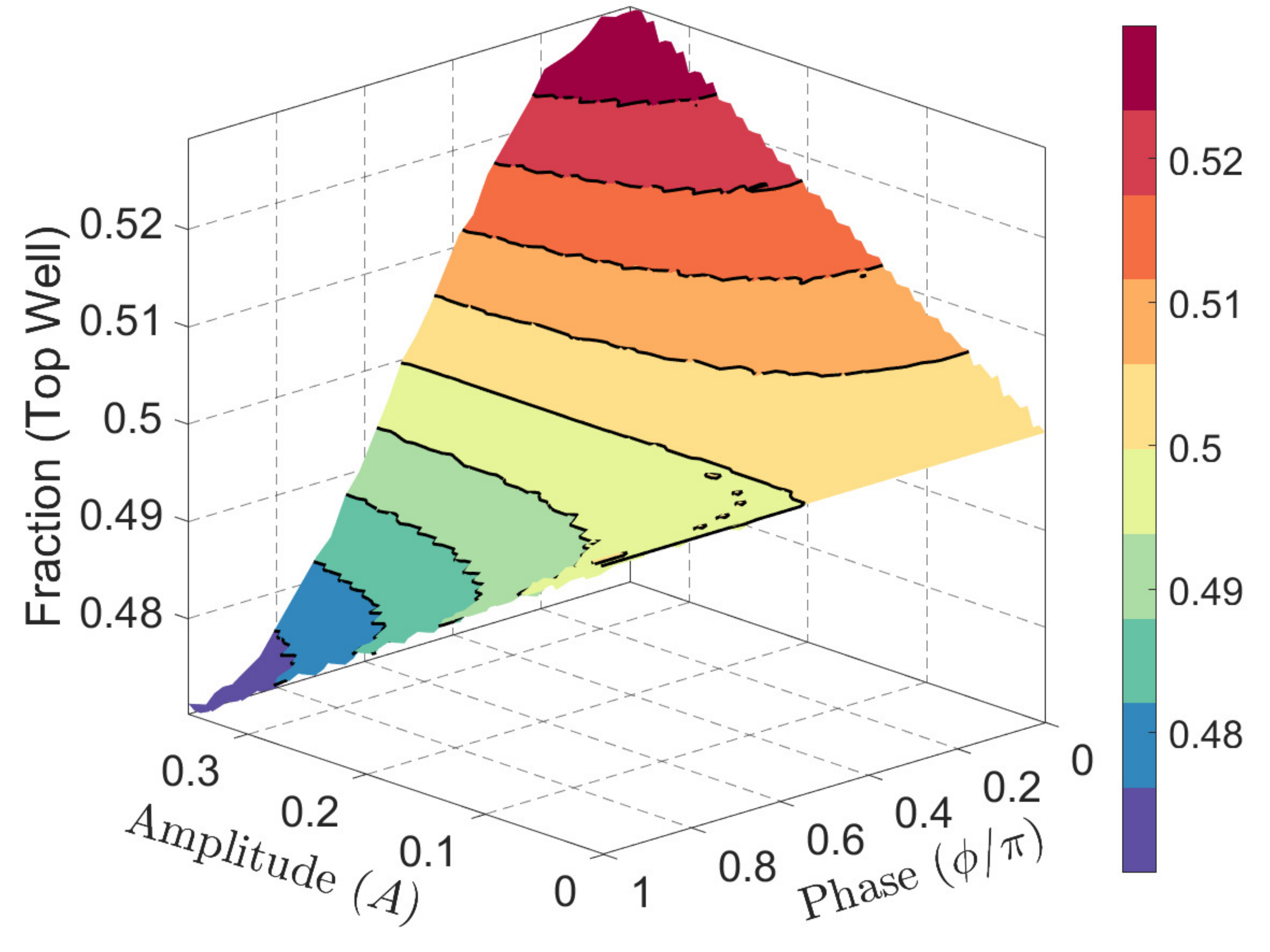}
		B)\includegraphics[scale=0.17]{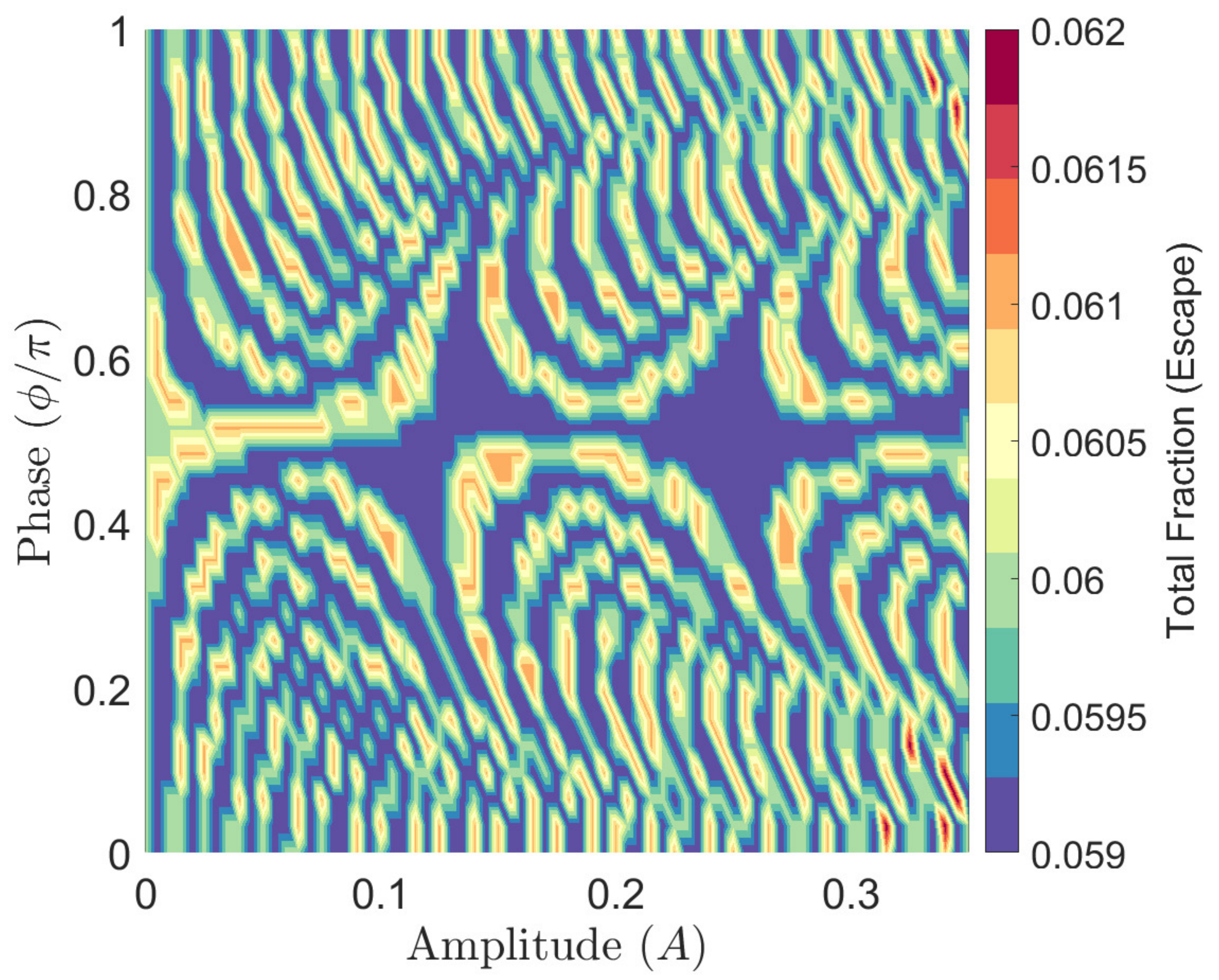}
	\end{center}
	\caption{Fraction of trajectories escaping the system or that enter either of the wells. The amplitude-phase parameter space has been calculated for a time dependent forcing characterized by an angular frequency $\omega = 8\pi$. A) Trajectories that enter the top well from all those that enter any of the wells. B) Fraction of trajectories that escape the system without entering either of the wells, calculated from the total number of trajectories of the initial ensemble.}
	\label{fig:paramSp_8pi}
\end{figure}

We finish this letter by briefly illustrating the response that the system displays when it is subjected to a forcing characterized by a very high frequency. In particular we take $\omega = 8 \pi$ and perform a similar analysis as the one we carried out above in the amplitude-phase parameter space. The expected behavior of the system is that, since the oscillations are very fast,  the trajectories generated from the underlying dynamics do not have enough time to respond to such rapid oscillations and, consequently,  they will follow paths very similar to the scenario where the system is fully time independent, i.e. $A = 0$, where the branching ratio is $1:1$. This is confirmed from the output of our calculations, which is included in Fig. \ref{fig:paramSp_8pi}. In panel A we can clearly see how the fraction of trajectories that go to the top well, calculated from all those that enter either well, give values very close to $0.5$ for the range of parameter values considered. Moreover, it is interesting to note that the fraction of trajectories that escape the system in this regime of high frequency forcing is indeed very small. This is depicted in panel B, where we see that the escape fraction exhibits interesting wave-like patterns in its distribution. We complement this analysis by doing a parameter study, but in this case fixing an amplitude of the forcing, $A = 0.15$, and sweeping through the angular frequency $\pi / 4 \leq \omega \leq 10 \pi$ and phase $0 \leq \phi \leq \pi$. From Fig. \ref{fig:paramSp_ampl} we can conclude that the fraction of trajectories that go to the top well vary from $0.35$ to approximately $0.83$ in the range of frequencies $0 \leq \omega \leq 2\pi$, but that it stabilizes to a value close to $0.5$ for frequencies higher than $\omega = 2\pi$. Moreover, panel B illustrates the escape fraction, and we can see how this quantity varies from $6\%$ to $12\%$ for the  interval of small frequencies $\pi / 4 \leq \omega \leq 2\pi$, while it drops below $6\%$ when the forcing frequency is increased beyond this interval.

\begin{figure}[htbp]
	\begin{center}
		A)\includegraphics[scale=0.19]{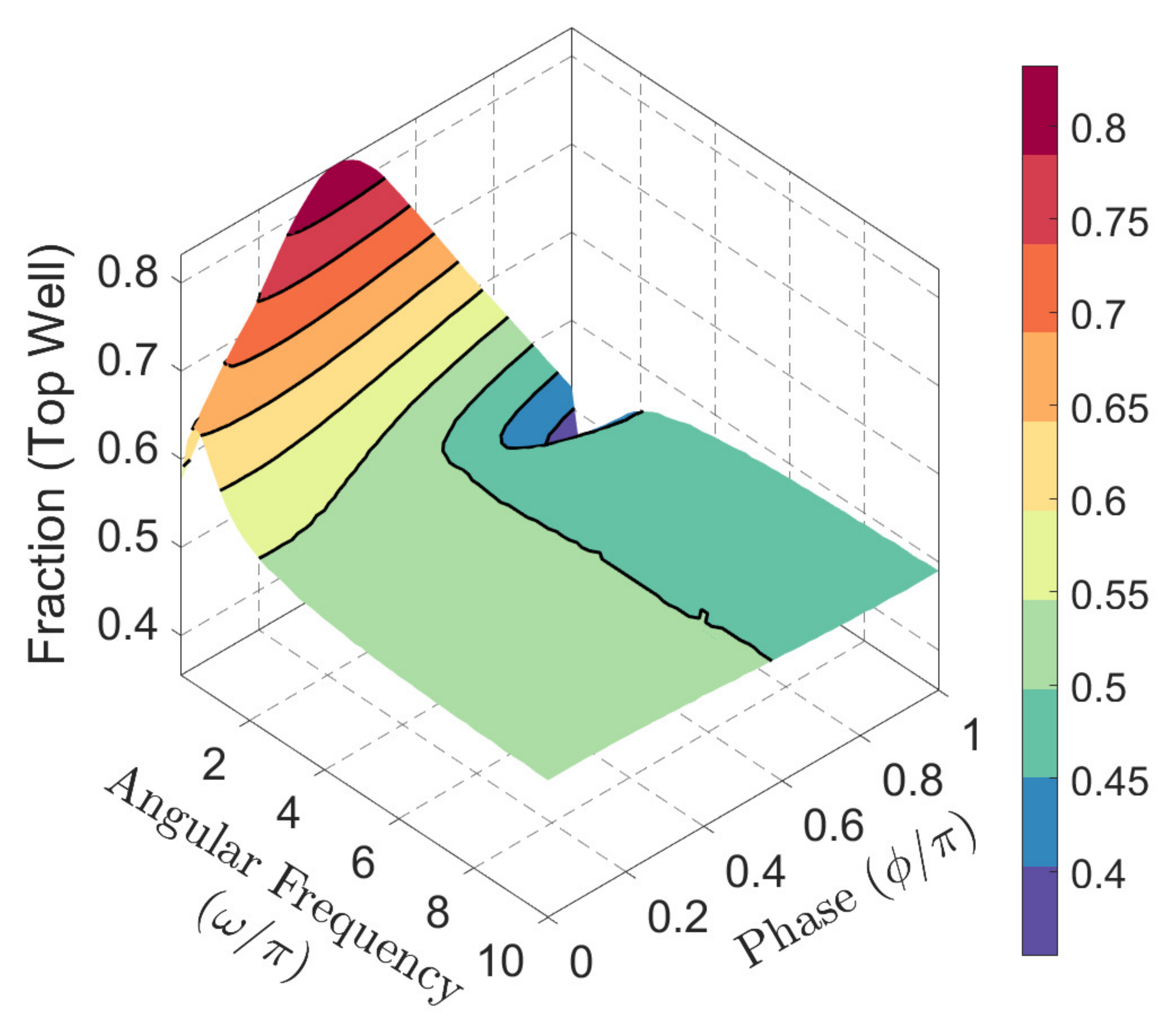}
		B)\includegraphics[scale=0.19]{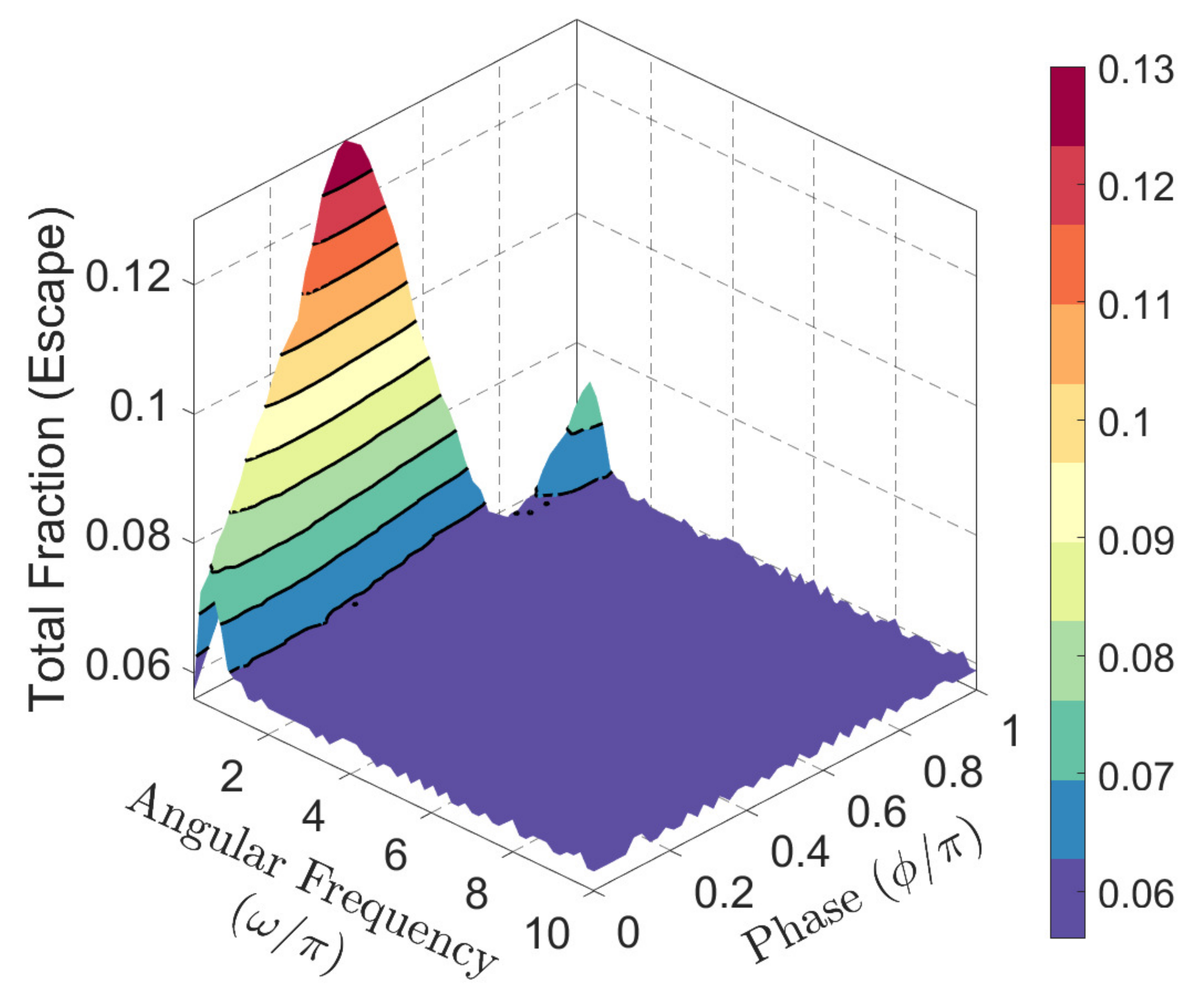}
	\end{center}
	\caption{Fraction of trajectories escaping the system or that enter either of the wells. The $(\omega,\phi)$-parameter space has been calculated for a time dependent forcing characterized by an amplitude $A = 0.15$. A) Trajectories that enter the top well from all those that enter any of the wells. B) Fraction of trajectories that escape the system without entering either of the wells, calculated from the total number of trajectories of the initial ensemble.}
	\label{fig:paramSp_ampl}
\end{figure}

\section{Summary and Outlook}
\label{sec:conc}

In this letter we have shown that chemical selectivity is exhibited on a symmetric  potential energy surface (PES) that features a valley-ridge inflection point (VRI) in the region between two sequential index-1 saddles, with one saddle having higher energy than the other when the system is subject to a time-periodic forcing. The forcing function depends on three parameters: amplitude, phase, and frequency. In the absence of forcing the branching ratio is $1:1$ as a consequence of the symmetry. However, we show that even though the PES is symmetric, the time-dependent forcing allows us to ``tune'' the branching ratio, essentially to any value that we desire by appropriate choice of parameters. We note the efforts that have been made in designing PESs that enforce a desired branching ratio \cite{campos2019designing}. Our results offer the possibility of a different approach to achieving a desired branching ratio.

Our analysis is based solely on trajectory calculations utilizing a numerical experiment guided by the geometry of the symmetric PES. In particular, we have not analysed the time-dependent phase space structure that governs selectivity in such a manner like it was analyzed for the symmetric, time-independent, PES in \cite{AGKW2020a}. This would be an intriguing analysis, and potentially very insightful, that we will leave for future investigations.

\section*{Acknowledgements}
The authors acknowledge the support of EPSRC Grant No. EP/P021123/1 and Office of Naval Research Grant No. N00014-01-1-0769.

\section*{Graphical Abstract}

\begin{figure}[htbp]
	\begin{center}
		\includegraphics[scale=0.19]{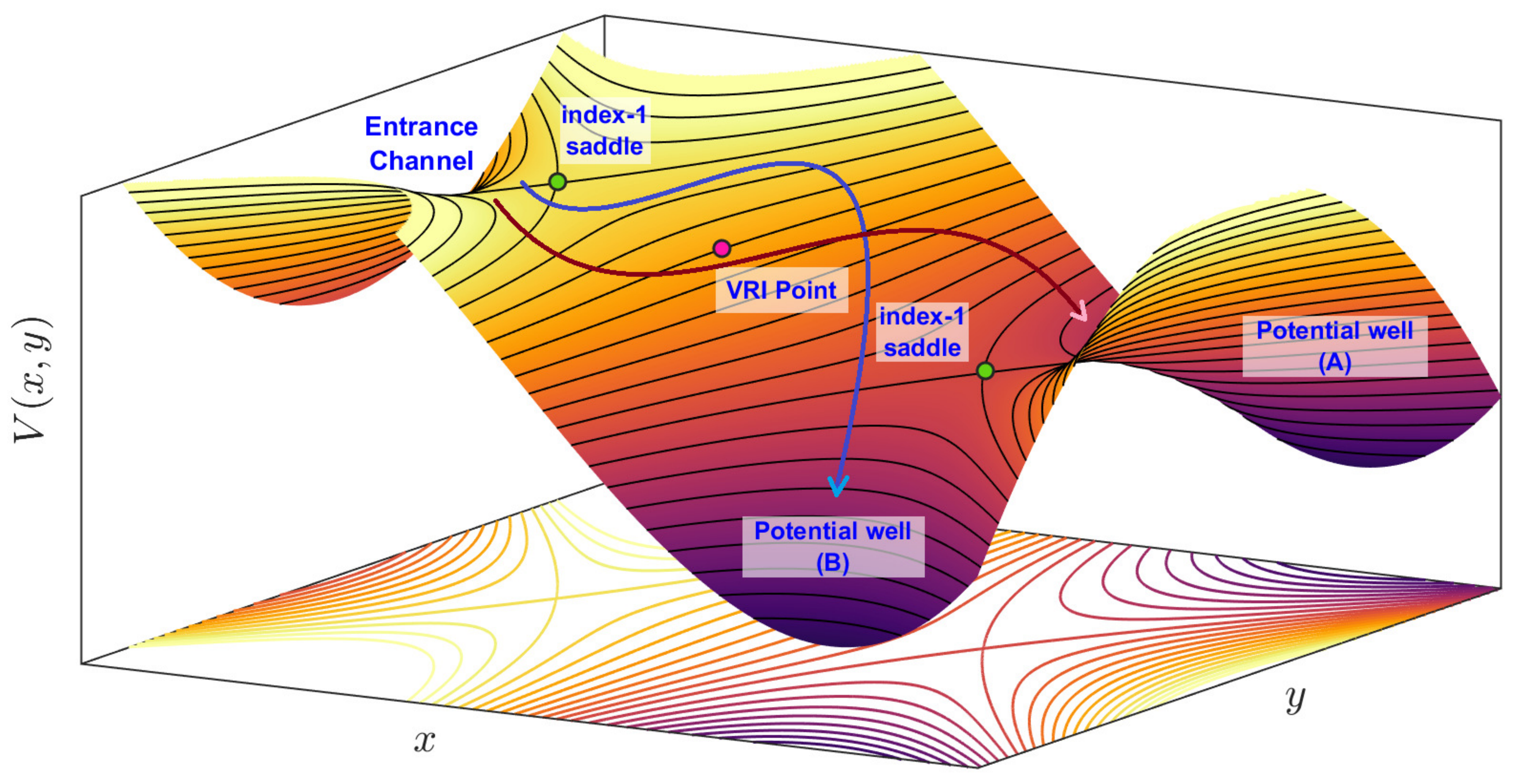}
	\end{center}
	\caption{Time-dependent forcing applied to the trajectories entering the potential energy surface through the upper channel governed by an index-1 saddle point. This interaction allows for the control of the selectivity mechanism in the chemical system.}
	\label{fig:paramSp_ampl}
\end{figure}

\bibliography{SNreac}

\end{document}